\documentclass[aps,onecolumn,preprint,superscriptaddress]{revtex4}
\usepackage{graphics}
\usepackage{graphicx}
\usepackage{amssymb}		
\usepackage{amsmath}
\usepackage{verbatim}
\usepackage{pstricks}
\newcommand{\bea}{\begin{eqnarray}}
\newcommand{\eea}{\end{eqnarray}}
\newcommand{\X}{{\cal X}}

\newcommand{\gsim}{\begin{array}{c}\sim\vspace{-26pt}\\> \end{array}}
\begin{document}

\begin{flushright}MCTP-09-14 \\
\end{flushright}
\vspace{0.2cm}

\title{Capture and Indirect Detection of Inelastic Dark Matter}

\author{Arjun Menon}
\affiliation{Michigan Center for Theoretical Physics (MCTP) \\
Department of Physics, University of Michigan, Ann Arbor, MI
48109
}

\author{Rob Morris}
\affiliation{Center for Cosmology and Particle Physics\\
Department of Physics, New York University, New York, NY
10003
}

\author{Aaron Pierce}
\affiliation{Michigan Center for Theoretical Physics (MCTP) \\
Department of Physics, University of Michigan, Ann Arbor, MI
48109
}

\author{Neal Weiner}
\affiliation{Center for Cosmology and Particle Physics\\
Department of Physics, New York University, New York, NY
10003
}

\date{\today}

\begin{abstract}
We compute the capture rate for Dark Matter in the Sun for models where the 
dominant interaction with nuclei is inelastic --- the Dark Matter up-scatters 
to a nearby dark ``partner'' state with a small splitting of order a 100 keV. 
Such models have been shown to be compatible with DAMA/LIBRA data, as 
well as data from all other direct detection experiments.  The kinematics of 
inelastic Dark Matter ensures that the dominant contribution to capture occurs 
from scattering off of iron.  We give a prediction for neutrino rates for current
and future neutrino telescopes based on the results from current direct 
detection experiments.  Current bounds from Super--Kamiokande and IceCube-22 
significantly constrain these models, assuming annihilations are into two-body 
Standard Model final states, such as $W^+W^-$, $t \bar t$, $b \bar b$ or $\tau^+ \tau^-$. Annihilations into first and second generation quarks and leptons are generally allowed, as are annihilations into new force carriers which decay dominantly into $e^+e^-$, $\mu^+ \mu^-$ and $\pi^+ \pi^-$.
\end{abstract}

\maketitle
\setcounter{equation}{0}
\section{Introduction}
Detection of dark matter is one of the most important - and challenging - tasks of modern astrophysics. A wide range of experiments have been undertaken to search for this dark matter directly and indirectly. Of the direct detection experiments, which hope to find the recoil of a Weakly Interacting Massive Particle (WIMP), the signature is the recoil of a nucleus after collision with a dark matter particle. The key task in these experiments is the distinction between the rare event (a WIMP scatter) and the common background (arising from natural radioactivity or cosmic rays).

One approach is to take advantage of the motion of the Earth relative to the rotation of the Milky Way spiral arms. 
As the Earth revolves around the sun, the WIMPs generally scatter more 
frequently in underground detectors when the orbit of the Earth moves it into 
the DM ``wind'' that arises from the galactic rotation from the (expectedly) non-rotating dark matter halo.  
This effect raises the possibility of searching for Dark Matter through  
an annual modulation 
signature~\cite{Drukier:1986tm,Freese:1987wu}. The DAMA/LIBRA experiment has 
released data that displays an unmistakable annual modulation, consistent with 
this Dark Matter interpretation~\cite{Bernabei:2008yi}.  If confirmed, this 
would give a valuable clue to the identity of the Dark Matter that makes up a 
quarter of the critical density of the Universe.  In the simplest 
models of Dark Matter, however, the DAMA/LIBRA signal is inconsistent with the the lack 
of observation in other low--background nuclear experiments, in particular
XENON~\cite{Angle:2007uj} and CDMS~\cite{Akerib:2005kh}. 

However, this conclusion does not apply to all Dark Matter models. It is 
possible to make the DAMA/LIBRA observation consistent with other direct dark 
matter detection experiments if what is being observed is actually inelastic 
scattering~\cite{TuckerSmith:2001hy} of the type
\begin{equation}
\chi + N \rightarrow \chi^{\ast} +N.
\end{equation}
In this process, the nuclear recoil spectra is sensitive to the mass difference 
$\delta \equiv M_{\chi^{\ast}} -M_{\chi}$.
Only WIMPs that have energies above the mass splitting can drive the 
up-scattering to the heavier state. For $\delta \sim 100$~keV, the recoil 
spectrum is sensitive to the target nucleus, raising the 
possibility that the scattering rates might differ drastically between 
experiments. This class of models has been shown to be consistent with both the 
recent DAMA/LIBRA data, as well as limits from the low-background 
experiments~\cite{WeinerKribs}, see also~\cite{MarchRussell:2008dy,Cui:2009xq}.  Inelastic Dark Matter could arise from mixed sneutrinos \cite{TuckerSmith:2001hy,TuckerSmith:2004jv,Thomas:2007bu} or from an SU(2)-doublet \cite{TuckerSmith:2004jv,Cui:2009xq}. It can also be incorporated into models that 
explain~\cite{ArkaniHamed:2008qn} the recent tantalizing excesses from the 
PAMELA~\cite{PAMELA} experiment, as well as an excess from INTEGRAL via the 
mechanism of Ref.~\cite{Finkbeiner:2007kk}.  Should this interpretation hold, 
both the positive result from DAMA and the null results from the other 
low-background experiments will have been crucial in pointing us towards the 
correct theory of Dark Matter. 

Indirect detection of Dark Matter through the annihilation into neutrinos can put strong constraints on Dark Matter models.
Determination of the neutrino rate requires knowledge of the Dark Matter 
annihilation rate in the Sun (or Earth) as well as the spectrum of the annihilation products.  
Typically, a neutrino signal is potentially observable when the annihilation 
rate is in equilibrium with the capture rate (see, e.g.,~\cite{JKG}).
This capture rate is determined by the interaction of the Dark Matter with 
nuclei.  Thus, the annihilation rate in the Sun is very sensitive to the 
scattering cross section off nuclei.  So, capture rates are tied closely to the
observed rate at DAMA/LIBRA, and neutrinos from the Sun are an important probe
of any mechanism that explains that signal.  Because of the gravitational 
potential of the Sun, WIMP capture in the Sun differs in important (but 
calculable) ways from the scattering off detectors in the Earth.

In this work, we first discuss the capture rate for inelastic Dark Matter (iDM)
in the Sun.  We then discuss the number of upward through-going muon-events 
that might be seen at detectors.   This depends on the final states of the annihilation of WIMPs. The annihilations produce Standard Model particles, which eventually decay to neutrinos, which then propagate from the Sun to the detectors on Earth.
A preliminary estimate of neutrino rates was made in~\cite{TuckerSmith:2001hy},
but here we extend the work, taking full account of the form factors in the 
scatterings, a proper treatment of the velocity distribution, as well as full propagation of the neutrinos from the center of the sun to the Earth. 

\section{Inelastic Capture rate for inelastic dark matter}
The central question in determining the rates of neutrinos is generally the WIMP capture rate on the sun. As WIMPs annihilate, one ultimately reaches equilibrium between capture and annihilation, i.e., $C_\odot = 2 \Gamma_{\odot}$. Thus, the capture sets the upper bound, and often the expected signal of neutrinos from the sun.
In this section we present analytical formulae for calculating the capture
rate of inelastic dark matter in the Sun. 

An inelastic WIMP of mass $m_\chi$ can
only scatter off a nucleus of mass $m_N$ if its energy $E \geq \delta (1+m_\chi/m_N)
$.
We will show a generalization of the results of
Refs.~\cite{Gould:1987ir} and~\cite{Gould:1991hx}, that allows for a discussion
of the capture of inelastic dark matter.

\subsection{Kinematics of Inelastic WIMP scattering}
The kinematics of inelastic scattering are quite different from that of elastic scattering. To have any possibility of scattering at all, the WIMP must satisfy the kinetic energy requirement
\bea
E_{min} \geq \frac{m_\chi v_{min}^2}{2} = \delta \left( 1 + \frac{m_\chi}{m_{N}}\right) 
\label{eqn:minE}.
\eea

The consequences of this change are purely kinematical. I.e., 
\bea
\frac{d\sigma_{elastic}}{d E_R} = \frac{d\sigma_{inelastic}}{dE_R},
\eea
but the allowed energy ranges for scattering can be quite different.
In the rest frame of the nucleus, conservation of energy and momentum of the
WIMP--nucleus scattering process implies a minimum
and maximum amount of energy loss in such a collision:
\bea
\Delta E_{min} &=& \frac{\delta}{\X_+} + 2 \frac{\X}{\X_+^2} \left(E -
\sqrt{E(E-\delta \X_+)} \right) \label{deltaEmin} \\
\Delta E_{max} &=& \frac{\delta}{\X_+} + 2 \frac{\X}{\X_+^2} \left(E +
\sqrt{E(E-\delta \X_+)} \right) \label{deltaEmax}
\eea
where
\bea
\X = \frac{m_\chi}{m_N} \;\;\;\;\;\;\; \X_{\pm} = \X \pm 1.
\eea
We use the parameter $\X$ to make connection with Ref.~\cite{Gould:1987ir}, which uses $\mu$. We use $\X$ to avoid confusion with the reduced mass parameter. This definition of $\X_{\pm}$ is twice that for $\mu$ in
Ref.~\cite{Gould:1987ir}.

As a consequence, the total cross section for inelastic scattering is 
\bea
\sigma_{inelastic} = \sqrt{1-\frac{2 \delta}{\mu v^2}} \sigma_{elastic}.
\eea
We will typically define our cross sections in terms of $\sigma_0$, which is the cross section {\em per nucleon} in the elastic limit.


\subsection{WIMP velocity distribution and flux}
To compute the scattering rate of the WIMPs on the Sun, we must characterize 
the distribution of WIMP velocities incident on the Sun. We assume that the 
WIMPs have an isothermal speed distribution in their rest frame. However, this 
frame need not coincide with the rest frame of the Sun.  In the Sun's rest 
frame, we have:
\bea
f(u) = \frac{1}{N(v_{esc})} \exp \left(-\frac{u_0^2}{\tilde{v}^2} \right),
\eea
where $\tilde{v}$ is the velocity dispersion, and
\bea
u_0^2 = u^2 + v_{Sun}^2 - 2 \vec{u} \cdot \vec{v}_s,
\eea
with $v_{Sun}$ is the speed of the Sun with respect to the WIMP rest frame.  The 
normalization factor, $N(v_{esc})$ is given by Ref.~\cite{Lewin:1995rx}:
\bea
N(v_{esc}) = (\pi \tilde{v}^2)^{3/2} \left({\rm erf}(y_m) - \frac{2}{\sqrt{
\pi}} y_m e^{-y_m^2} \right),
\eea
where $y_m = v_{max}/\tilde{v}$, and $v_{esc}$ is the escape velocity.  In the limit of $y_{m} \rightarrow \infty$ this expression simplifies to $N(\infty)= (\pi \tilde{v}^2)^{3/2}$.

We now wish to compute the flux distribution of WIMPs incident on the Sun.  We begin by considering an imaginary surface bounding a region of radius $R$ about
the Sun, which is large enough so that the gravitational field due
to the Sun is negligible. For an infinitesimal region of area $R^2 \sin \theta
d\theta d\phi$ we choose a local coordinate system (see Fig.~1) so that
\bea
\vec{u} &=& u \cos \alpha \hat{r} + u \sin \alpha \cos \beta \hat{\theta} +
u \sin \alpha \sin \beta \hat{\phi} \\
\Rightarrow \vec{u} \cdot \vec{v}_s &=& u v_{Sun} (\cos \alpha \cos \theta - \sin
\alpha \cos \beta \sin \theta).
\eea
\begin{figure}
\includegraphics{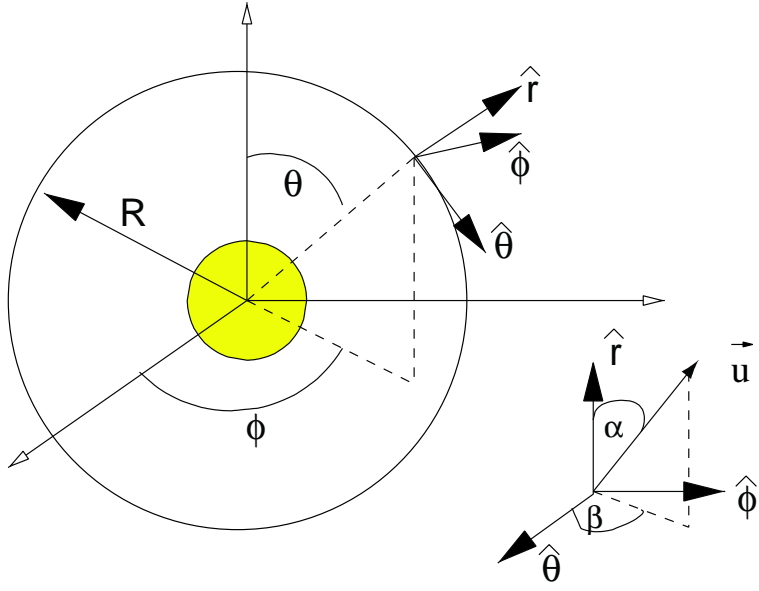}
\caption{Coordinate system used in the derivation of the capture rate.  The inset is the local coordinate system on the surface of the sphere, which defines the angles $\alpha$ and $\beta$.}
 \label{fig:solarcoords}
\end{figure}
Therefore the flux of WIMPs with velocity in the range $(u,u+du),(\alpha ,
\alpha + d\alpha),(\beta, \beta+d\beta)$ through the infinitesimal area
$R^2 \sin \theta d\theta d\phi$ is
\bea
d\mathcal{F} &=& \exp \left(-\frac{u^2 + v_{Sun}^2 - 2 u v_{Sun} (\cos \alpha
\cos \theta - \sin \alpha \sin \theta \cos \beta)}{\tilde{v}^2} \right)
\times \nonumber \\
& & R^2 \sin \theta d\theta d\phi \left(-\vec{u}\cdot \hat{r}
\right) \frac{n_{DM}}{N(v_{esc})} u^2 du d(\cos \alpha) d\beta.
\eea
If we make the approximation that $v_{esc} \rightarrow \infty$, then the analysis simplifies considerably.  As we will see, we do not expect this assumption to drastically affect the results.  With this assumption is possible to do a straightforward integration over $\beta$  (and $\phi$) 
\bea
d\mathcal{F}&=& -\exp \left(-\frac{\left(u^2 + v_{Sun}^2 - 2 u v_{Sun} (\cos \alpha
\cos \theta) \right)}{\tilde{v}^2} \right) I_0\left(\frac{u v_{Sun}}{\bar{v}^2}
\sin \alpha \sin \theta \right) \times \nonumber \\
& & R^2 \sin \theta d\theta \frac{2\pi^2 n_{DM}}{N(v_{\infty})} u^3 du
d(\cos^2 \alpha).
\eea
Here, $I_0$ is the modified Bessel function, 
and $n_{DM} \equiv \rho_{DM}/m_\chi$ is the WIMP number density. We now 
change variables to angular momentum per unit mass $\equiv J$. Therefore
\bea
\cos \alpha = \sqrt{1 - \frac{J^2}{u^2 R^2}},
\eea
and the total flux of WIMPs becomes
\bea
d\mathcal{F} &=&\exp \left(-\frac{u^2 + v_{Sun}^2 - 2 u v_{Sun} \cos \theta \sqrt{1 -
\frac{J^2}{u^2 R^2}}}{\tilde{v}^2} \right) I_0\left(\frac{Jv_{Sun}}{R\tilde{v}^2}
\sin \theta \right) \times \nonumber \\
& & \sin \theta d\theta \frac{2\pi^2 n_{DM}}{N(\infty)} u du dJ^2 \\
&=&\exp \left(-\frac{u^2 + v_{Sun}^2 - 2 u v_{Sun} \cos \theta}{\tilde{v}^2} \right)
 d(\cos \theta) \frac{2\pi^2 n_{DM}}{N(\infty)} u du dJ^2,
\label{dflux}
\eea
where in the last line we take the $R\to \infty$ limit. Therefore as
expected the flux of WIMPs in the forward direction is greater than that in
the direction opposite to the velocity of the Sun.

\subsection{Differential capture rate}

Following the calculation in Ref.~\cite{Gould:1987ir}, the WIMP
velocity at the some $r$ inside the Sun is
\bea
w^2 = u^2 + v^2(r)
\eea
and the probability for the WIMP to scatter inside a shell of thickness $dr$ at
radius $r$ is
\bea
& & \Omega_v^-(w) \frac{dl}{w} \\
&=& \Omega_v^-(w) \frac{2}{w} dr \left(1-\frac{J^2}{r^2 w^2} \right)^{-1/2}
\Theta (rw-J)
\eea
Therefore the differential capture rate is a product of the flux times the
probability to scatter:
\begin{eqnarray}
dC_{\bigodot} &=& \Omega_v^-(w) 4r^2 w \exp \left(-\frac{u^2 + v_{Sun}^2 - 2 u v_{Sun} \cos
\theta}{\tilde{v}^2} \right) d\cos \theta dr \nonumber \\
&\times& \frac{2\pi^2 n_{DM} u }{N(\infty)} du \; 
\Theta(\frac{1}{2} m w^2 - \delta \X_+).
\end{eqnarray}
To arrive at this expression, we performed the $J^2$ integral from $0$ to $r^2 
w^2$.  The step function $\Theta(\frac{1}{2} m_\chi w^2 - \delta \X_+)$ imposes the
constraint that the energy of the
WIMP is sufficient for the collision to occur. Therefore we find
\bea
\frac{dC_{\bigodot}}{dV} &=& \int_{-1}^{1} dx \int_0^{\infty } f(u,x)du
\frac{w}{u} \Omega_v^-(w) \; \Theta(\frac{1}{2} m_\chi w^2 - \delta \X_+)
\label{dcdv}
\eea
where
\bea
f(u,x)du &=& \frac{2\pi n_{DM}}{N(\infty)}\exp \left(-\frac{u^2 + v_{Sun}^2 - 2
u v_{Sun} x}{\tilde{v}^2} \right) u^2 du \\
&=& \frac{4 \pi n_{DM} \tilde{v}^2}{N(\infty) u v_{Sun}} \exp \left(-\frac{u^{2} + v_{Sun}^{2}}{\tilde{v}^{2}}\right) \sinh \left(\frac{2 u v_{Sun}}{\tilde{v}^{2}}\right) u^{2} du.
\eea
where in the last line we have performed the integral over $x$. 
Note that if we had not taken the $v_{esc} \rightarrow \infty$ limit, the upper limit of integration on the $u$ integral would have been both $\alpha$ and $\beta$ dependent, greatly complicating the analysis.
Generalizing the result in Eqn.~(A5) of Ref.~\cite{Gould:1987ir} we have 
\bea
\Omega_v^-(w) &=&\frac{ \sigma n w}{E_{max}^{elastic}} \int_{\Delta E_{min}}^{\Delta E_{max}} d(\Delta E)
F^2(\Delta E)
\Theta(\Delta E-E_{\infty}), \label{omegaminus}
\eea
where $E=mw^2/2$, $E_{\infty}=m_\chi u^2/2$, and $E_{max}^{elastic} = 4 \mu^2 v^2/m_N$ is the maximum nuclear recoil energy in the elastic case. $F^2(\Delta E)$ is the form factor that 
accounts for the fact that at sufficiently high momentum transfer, scattering 
from the nucleus is no longer coherent.  We have also defined
\bea
\sigma \equiv \sigma_0 \frac{m_\chi^2 m_N^2}{m_n^2 (m_\chi+m_N)^2} \frac{(f_p Z + f_n (A-Z))^2}{f_n^2},
\label{eqn:sigdef}
\eea
where $m_n$ is the nucleon mass and $\sigma_0$ is the WIMP-nucleon scattering 
cross-section. $f_{p,n}$ are the relative proton and neutron couplings, which we take to be equal. 
The results for the differential capture rate in 
Eqns.~(\ref{dcdv})--(\ref{eqn:sigdef}) represent the main analytic result of 
this paper.  

\subsection{Inelastic dark matter capture and astrophysical
uncertainties}
In this section we discuss the implications of Eqn.~(\ref{dcdv}) and
the uncertainty in the capture rate due to the astrophysical uncertainties in
$v_{Sun}$, $\tilde{v}$ and the distribution of heavy nuclei in the
Sun. We also consider the effect of changes in
$F^2(\Delta E)$ in Eqn.~(\ref{omegaminus}). The changes in the capture rate 
that result from varying these astrophysical and nuclear parameters give 
us an estimate of some of the systematic uncertainties in our calculation and 
an idea of the robustness of the current limits we discuss in 
Sec.~\ref{sec:presentlimits}. As a reference point, we use the values 
$\rho_{DM}=0.3$~GeV/cm$^{3}$ and $\sigma_0 = 10^{-40}$~cm$^2$ throughout this 
section: a change in these values will just rescale the total capture rate.
Also, as reminder, this value of $\sigma_0$ is chosen because the inelastic
cross-section needed to make the DAMA/LIBRA results compatible with direct 
detection experiments is of the same order of magnitude.

\begin{figure}
\begin{center}
\resizebox{9.cm}{!}{\includegraphics{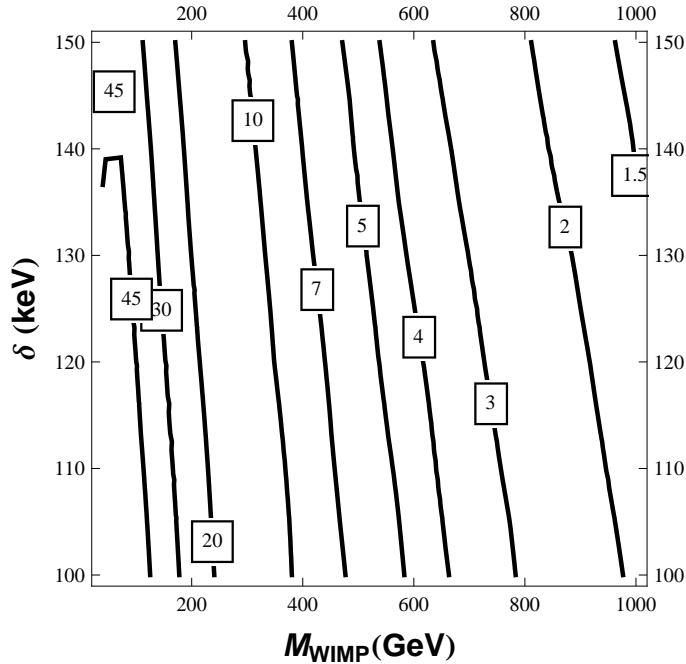}}
\end{center}
\caption{Contour plot of  $C_{\bigodot}/10^{23} s^{-1}$ in the $M_{WIMP}-\delta$ plane, where $v_{Sun} = 250$~km/s, $\tilde{v} = 250
$~km/s, $\rho_{DM} = 0.3$~GeV/cm$^3$ and $\sigma_0=10^{-40}$cm$^2$.}
\label{mdcapplot:fig}
\end{figure}

\begin{figure}
\begin{center}
\resizebox{8.cm}{!}{\includegraphics{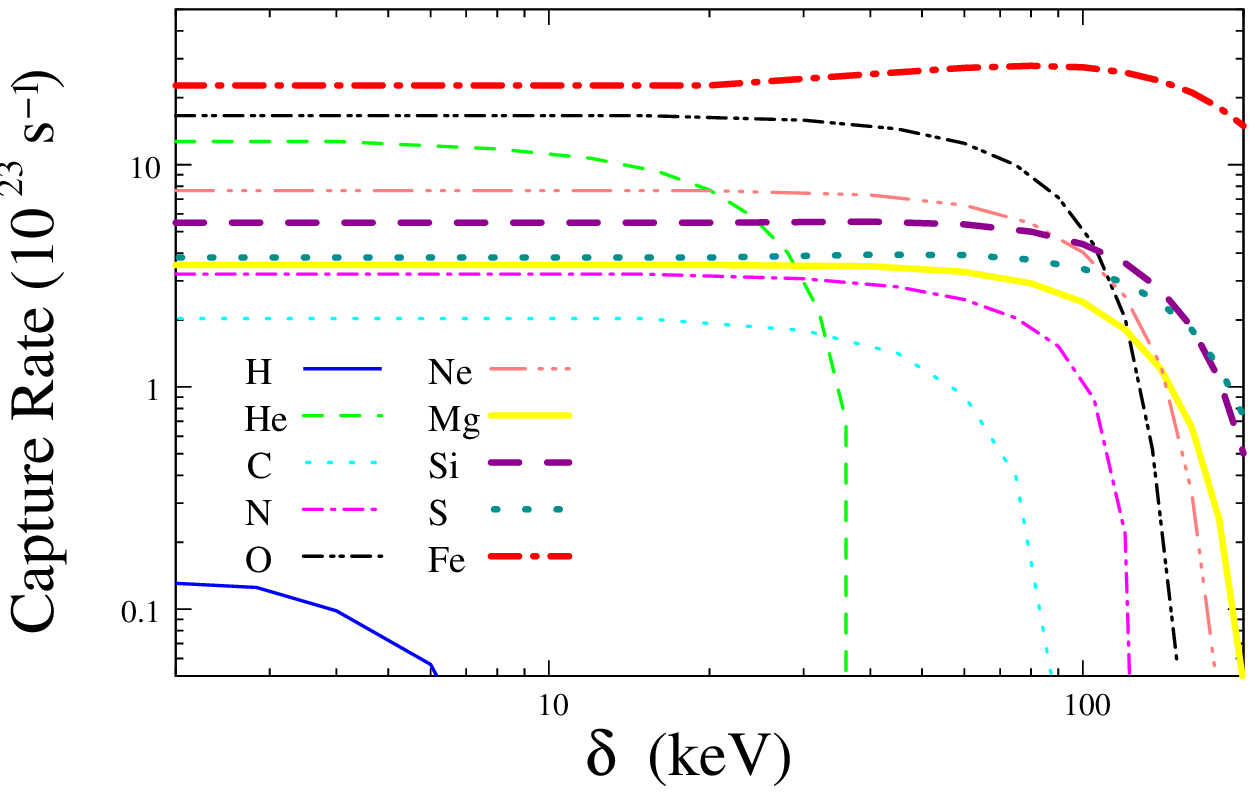}}
\resizebox{7.2cm}{!}{\includegraphics{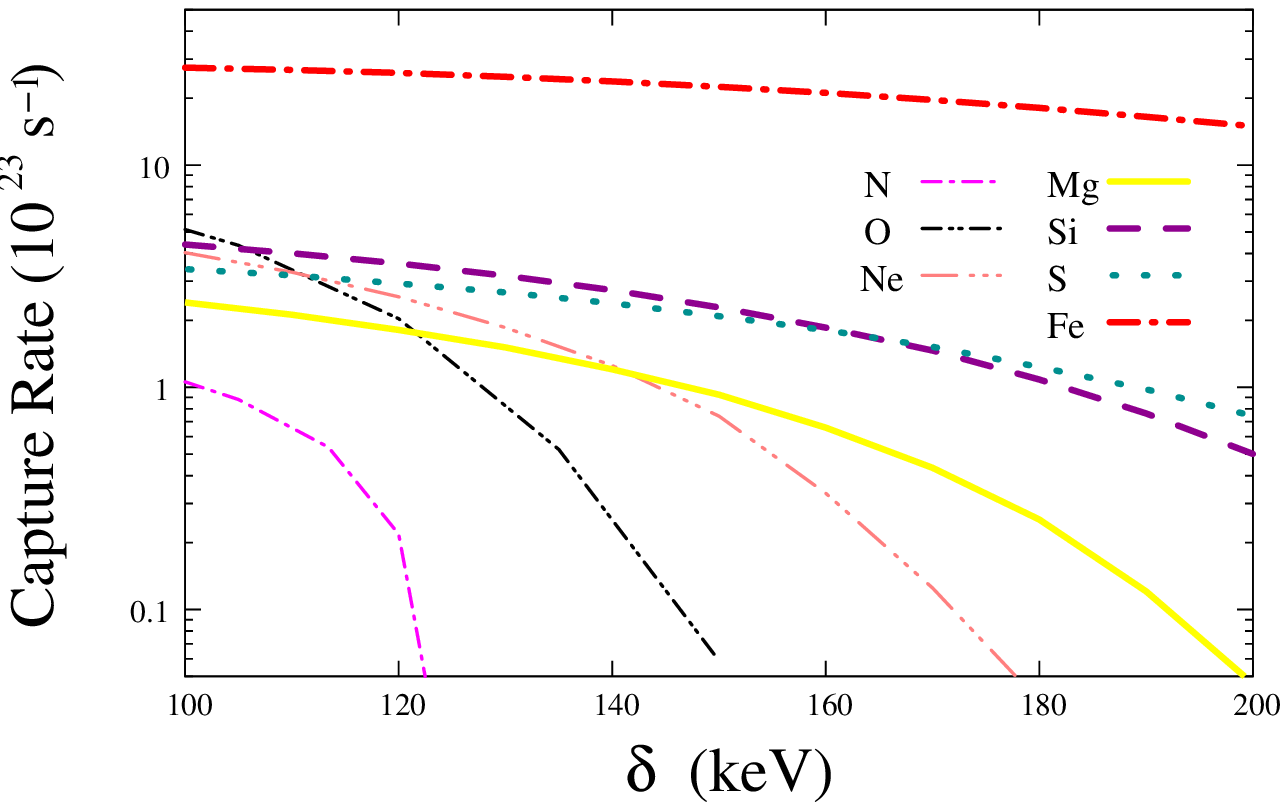}}
\end{center}
\caption{(left) Capture rate of a $100$~GeV WIMP due the different species of
nuclei in the Sun, assuming the standard solar profile of
Ref.~\cite{Bahcall:2004pz,PenaGaray:2008qe} and the same values of $v_{Sun},
\tilde{v},
n_{DM}$ and $\sigma_0$ as in Fig.~\ref{mdcapplot:fig}.(right) Expanded view of (a) for
the region of $\delta > 100$~keV.
}
\label{elem:fig}
\end{figure}

Before exploring the effects of these astrophysical uncertainties, we establish
a baseline in Fig.~\ref{mdcapplot:fig}.  There we show the variation of the 
capture rate in the $m_{WIMP}-\delta$ plane for the values $v_{Sun} = 250$~km/s,
and $\tilde{v} = 250$~km/s.  We have also used the Helm 
form factor from Ref.~\cite{Lewin:1995rx} for $F^2(\Delta E)$. As in the case 
of elastic dark matter, the capture rate decreases with increasing WIMP mass. 
This 
is in part due to a simple decrease in the number density of WIMPs.  However, 
there are two additional effects: it is more difficult for heavy WIMPs to 
lose energy in collisions with the relatively light nuclei in the Sun, and as 
the mass of the WIMP increases, it becomes more difficult to satisfy the 
minimum scattering energy condition of Eqn.~(\ref{eqn:minE}). 
The minimum scattering energy condition in Eqn.~(\ref{eqn:minE}) leads to a suppression of the capture rate due to lighter nuclei, as shown in Fig.~\ref{elem:fig}.

To focus on the effect that the inelasticity has on WIMP capture, in 
Fig.~\ref{elem:fig} we show the capture rate of a $100$~GeV WIMP on each species of nuclei in the Sun.  We assume the same values of $v_{Sun},
\tilde{v},n_{DM}$ and $\sigma_0$ as in Fig.~\ref{mdcapplot:fig}.
We see
that for $\delta \gsim 30$~keV, the scattering of WIMPs off hydrogen and
helium is highly suppressed, and in the range $100 \mbox{ keV} < \delta < 150
$~keV the capture due to scattering off iron dominates the other elements by a 
factor of$~4$.  As a reminder, this range of values $100 \mbox{ keV} < \delta < 150$~keV
provides the best fit to the current DAMA data. 
The importance of the heavy elements for the capture rate is not unique 
to the inelastic case: for both elastic and inelastic 
(spin-independent) models scattering off the heaviest elements can dominate. (This is not true of spin-dependent scattering.)
This is perhaps counter--intuitive, as the abundances of the heavy elements are 
substantially less than hydrogen and helium.  However, these tiny abundances 
are compensated by the fact that the (coherent) spin-independent couplings 
scale as the square of the atomic number. In fact, when one accounts for 
additional kinematic factors, for large Dark Matter masses the 
spin-independent cross-section scales as the 
fourth power of the atomic mass number. These $A^{4}$ factors can more than 
overcome the relative scarcity of heavy elements.  Some 
reviews in the literature dangerously neglect the contributions of these heavy 
elements for simplicity. 

\begin{figure}
\includegraphics{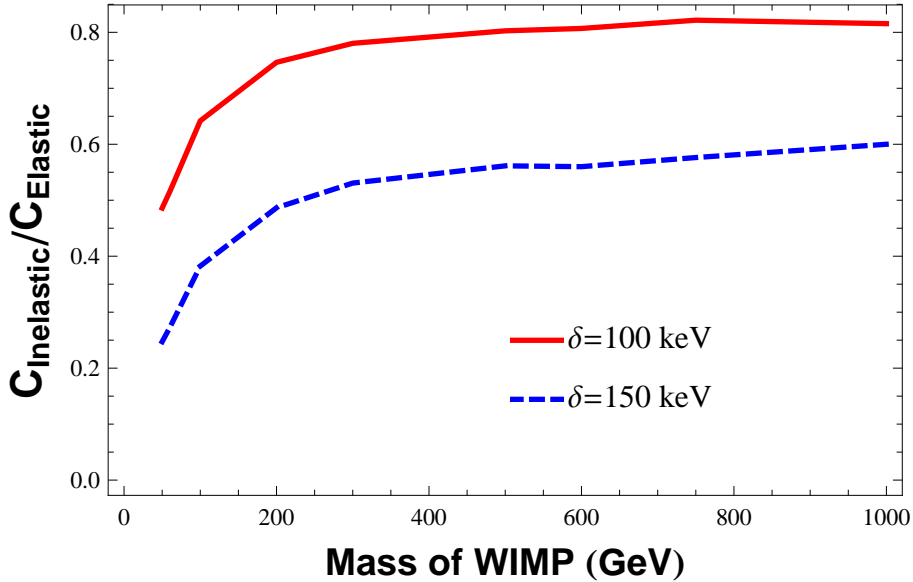}
\caption{The suppression of the capture rate in the inelastic case relative to the elastic scattering case with comparable $\sigma_0$ shown as a function of $M_{WIMP}$ for two choices of $\delta=100,150$ keV.}
\label{fig:elasticVinelastic}
\end{figure}

Depending on the choice of $\delta$, there is actually a slight enhancement of 
capture due to iron relative to the elastic case.
The reason is that form factor suppression is smaller for the inelastic case 
compared to the elastic one: for the same loss in WIMP energy $\Delta E$ the
recoil energy of the nucleus is $E_R = \Delta E - \delta$ for the inelastic 
scenario compared to $E_R = \Delta E$ for WIMPs that scatter elastically.
The net effect of inelastic scattering compared to the elastic case is shown 
in Fig.~\ref{fig:elasticVinelastic}. 

\begin{figure}
\begin{center}
\resizebox{8.cm}{!}{\includegraphics{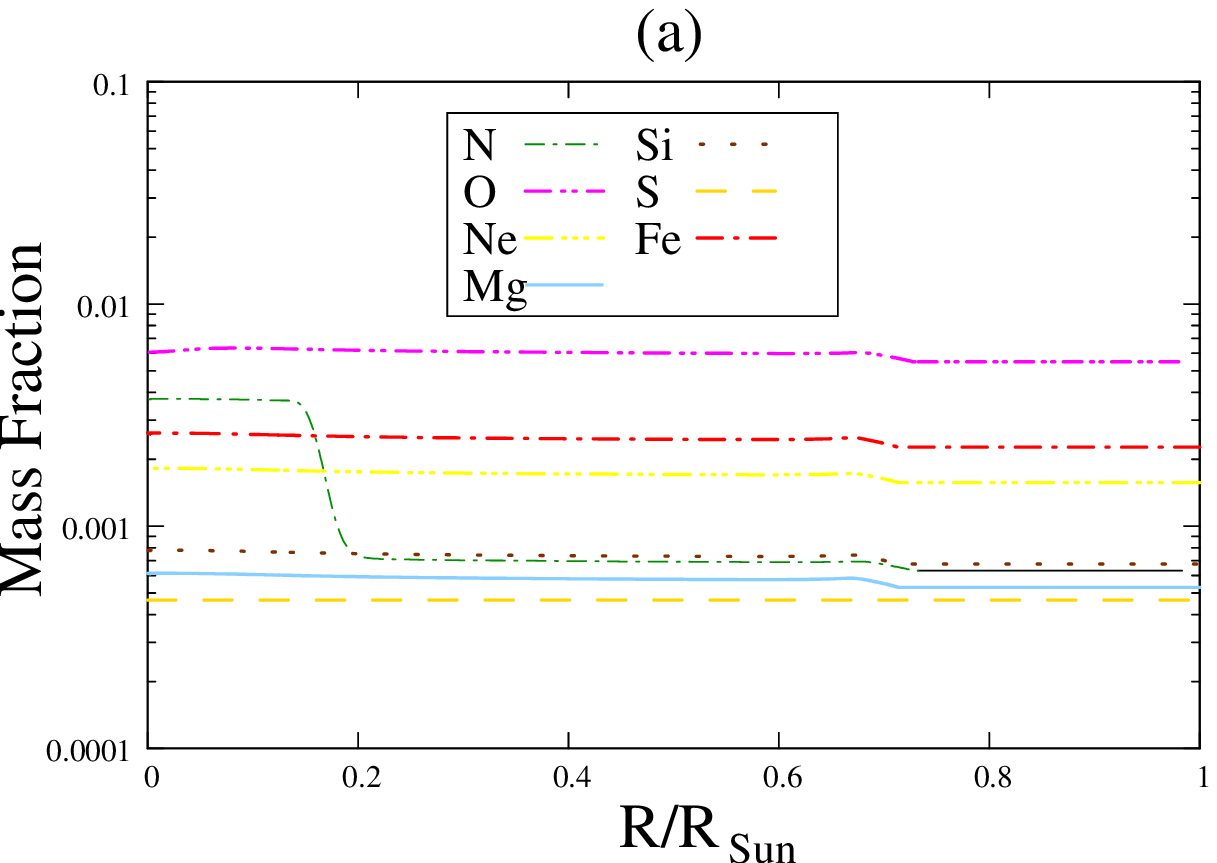}}
\resizebox{8.cm}{!}{\includegraphics{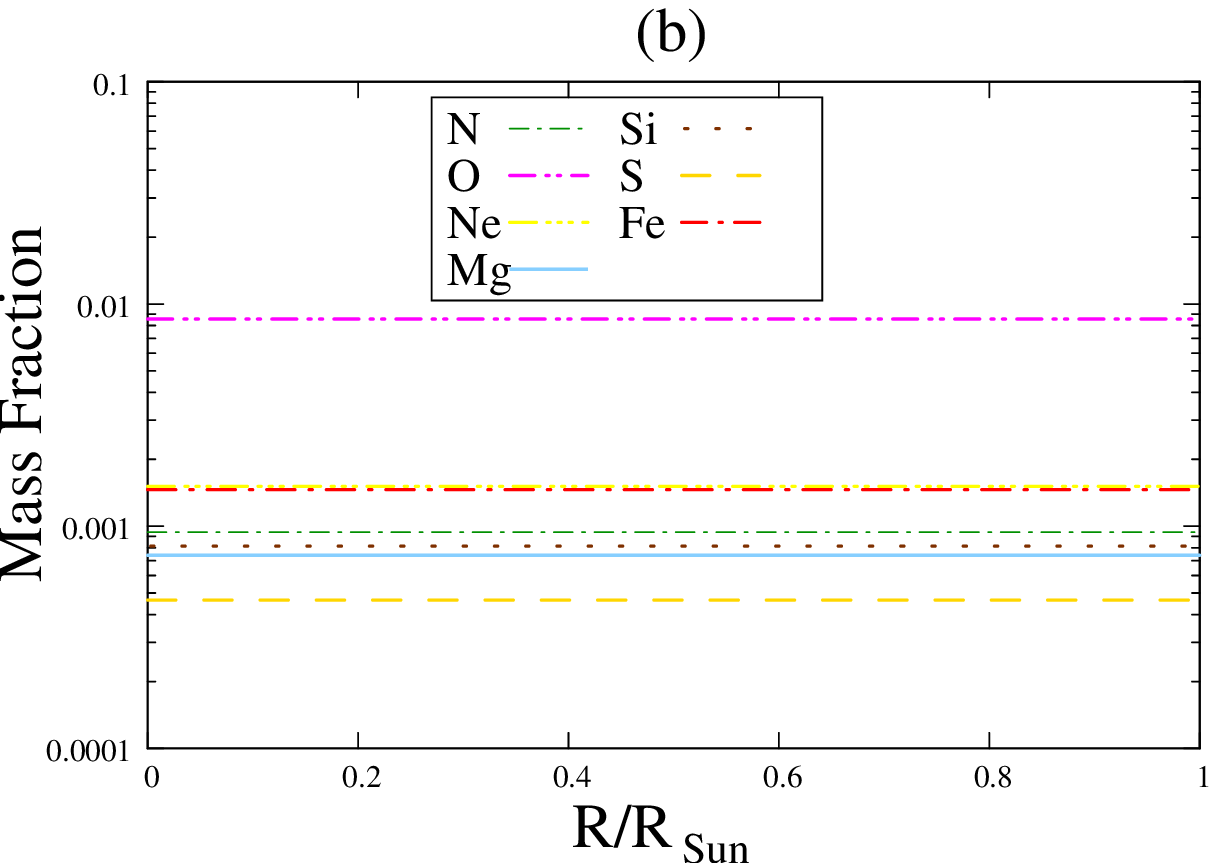}}
\resizebox{8.cm}{!}{\includegraphics{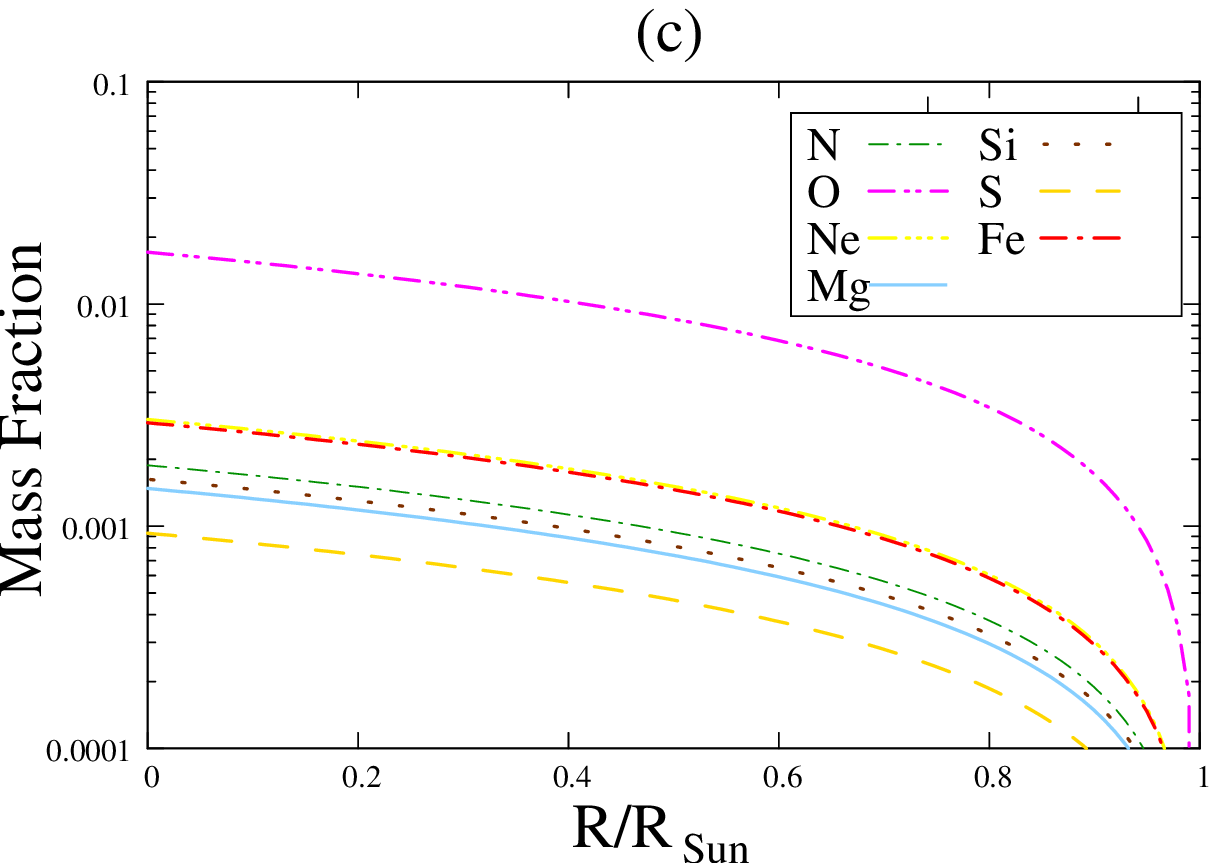}}
\resizebox{8.cm}{!}{\includegraphics{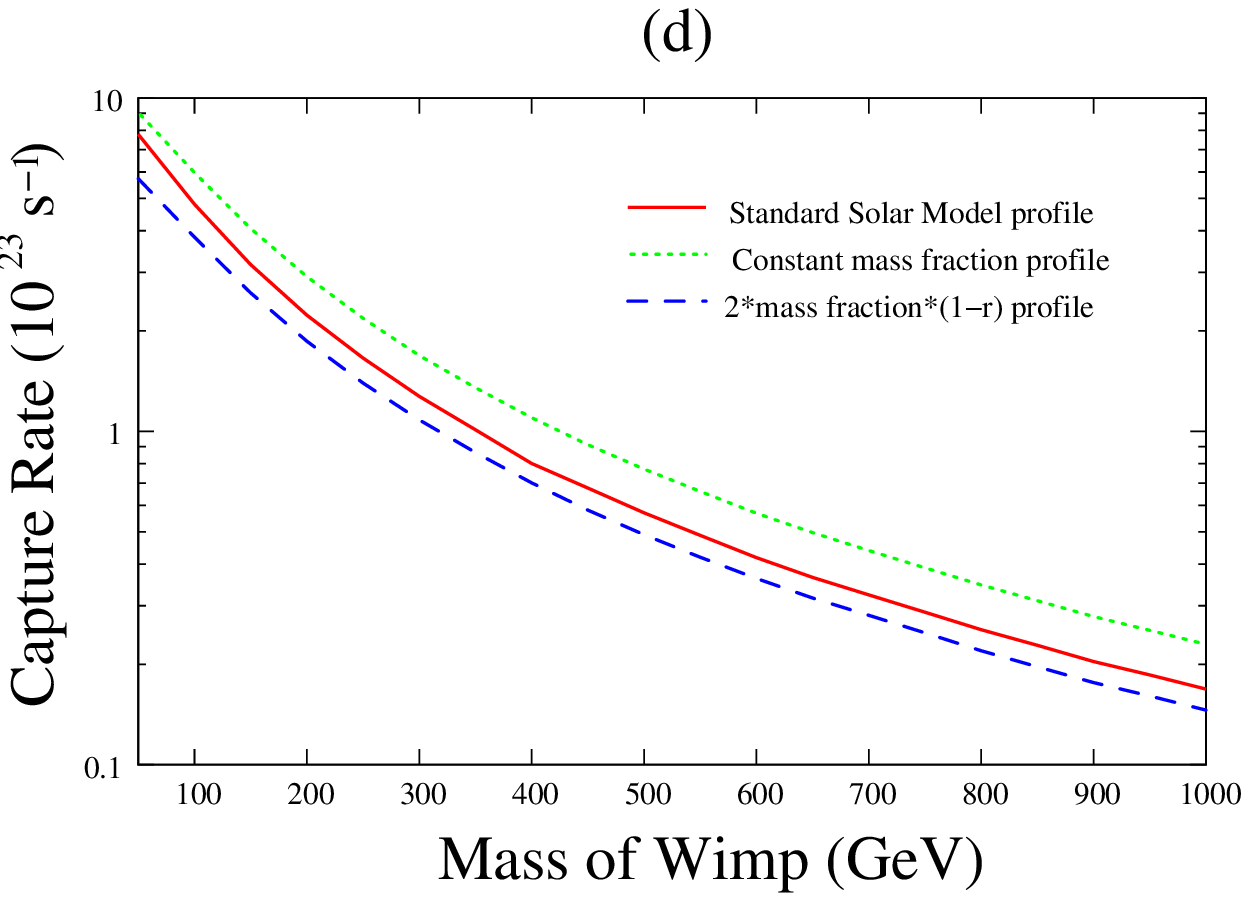}}
\end{center}
\caption{(a) Standard Solar profile from Ref.~\cite{Bahcall:2004pz,PenaGaray:2008qe}. (b) 
Constant profile. (c) Linearly decreasing profile. (d) Effect of changing the 
profile of the heavy elements in the Sun on the
capture rate of inelastic WIMPs for $\delta = 100$~keV with the remaining
inputs the same as in Fig.~\ref{mdcapplot:fig}.}
\label{profile:fig}
\end{figure}

We now turn to sources of uncertainty in the capture rate.  The importance of 
heavy nuclei in the capture of inelastic WIMPs implies a sensitivity to the abundance 
and distribution of these nuclei in the Sun, which is somewhat uncertain. To 
quantify this dependence on the distribution of heavy nuclei in the Sun we 
consider two variations to the Standard Solar Model of 
Ref.~\cite{Bahcall:2004pz,PenaGaray:2008qe}: a uniform distribution as a function of radius for 
all heavy nuclei and a linearly decreasing distribution with the same average 
number density as the uniform one (which we view as somewhat of an extreme 
case). We show these distributions, along with the resulting capture rates in 
Fig.~\ref{profile:fig}.  Again, we have set $\delta =100$~keV and 
values of $v_{Sun},\tilde{v},n_{DM}$ and $\sigma_0$ that are the same as
those in Fig.~\ref{mdcapplot:fig}. From Fig.~\ref{profile:fig} we see that if 
the number density of heavy nuclei in the center of the Sun is increased the 
overall capture rate can be significantly increased.  Also, the mass fraction 
of iron in the Sun will directly affect the capture rate.

\begin{figure}
\begin{center}
\resizebox{12.cm}{!}{\includegraphics{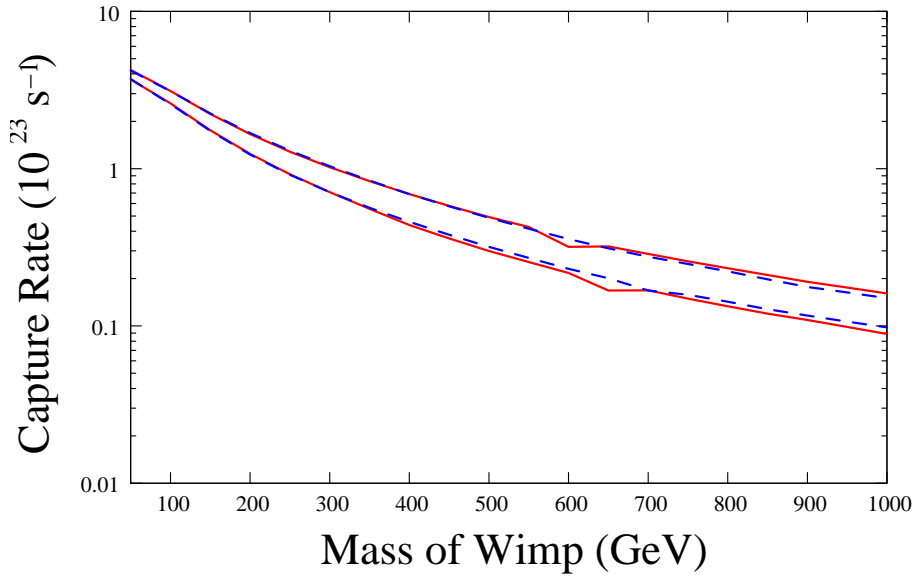}}
\end{center}
\caption{Dependence of the total capture rate for $\delta = 150$~keV  on
$v_{Sun}$ and $\tilde{v}$. 
The red (solid) curves bound the region swept out by $C_{\bigodot}$ for the range of $200 \mbox{ km/s} < v_{Sun} < 300 \mbox{ km/s}$ and the blue (dashed) band corresponds to the
values of $C_{\bigodot}$ for the range of $200 \mbox{ km/s} < \tilde{v} < 300 
\mbox{ km/s}$.}
\label{vmvsvb:fig}
\end{figure}

The threshold for scattering in Eqn.~(\ref{eqn:minE}) also suggests that the 
capture
rate is sensitive to variations in the astrophysical inputs $v_{Sun}$
and $\tilde{v}$, as well as perhaps the escape velocity.  First, we address the issue of the finite escape velocity.  We do not expect the capture rate to be very sensitive to this 
quantity.  This is because the particles that are most easily captured are {\it
 not} those particles in the Boltzmann tail.  These particles have the most 
energy and are harder to capture.  This is in strict contrast to scattering at 
DAMA/LIBRA, where only the most energetic particles in the Dark Matter distribution 
are able to drive the inelastic transition, simply because these particles do 
not have the benefit of mining the gravitiational potential energy of the Sun.  
(The intuition that the result is insensitive to $v_{esc}$ was crudely confirmed by varying the (large) limit of the $u$ integral when performing the numerical integration, and seeing that the capture rate was insensitive.) In Fig.~\ref{vmvsvb:fig} we show the predicted range of values
$200 \mbox{ km/s} \leq v_{Sun} \leq 300$~km/s with a central value of
$250$~km/s in the red (solid) band and $200 \mbox{ km/s} \leq
\tilde{v} \leq 300$~km/s with a central value of $250$~km/s in the blue 
(dashed) band, where we have used
the Standard Solar Model heavy nuclei number densities, $\rho_{DM} = 0.3
$~GeV/cm$^3$and $\sigma_0=10^{-40}$cm$^2$.  If one assumes no co-rotation of 
the WIMP halo,  then the velocity of the Sun relative to the WIMP halo is 
determined as $v_{rot} + v_{peculiar}$ with $|v_{peculiar}| \sim 20
$~km/s~\cite{SolarVel}.  Recent measurements place $v_{rot}=250$~km/s.  We take
 this as a central value, but allow a range values to account for the 
possibility of 
errors in its determination and some co/counter-rotation.  If one further 
assumes that the halo is virialized then $\tilde{v} = v_{rot}$, but we allow 
$\tilde{v}$ and $v_{Sun}$ to vary independently in the plot. 


The variation in the capture rate illustrated in Fig.~\ref{vmvsvb:fig} is 
mainly due to the fact that mostly only slow moving WIMPs can lose sufficient
energy to be captured. This effect dominates that of the WIMP minimum 
energy scattering condition in Eqn.~(\ref{eqn:minE}). Therefore increasing 
$\tilde{v}$ has the effect of decreasing the capture rate.
This result is the same as that observed in Refs.~\cite{Gould:1987ir} 
and~\cite{Gould:1991hx}.   Even varying over this generous range, only a factor of two change in the capture rate is observed.



Finally, we note the importance of form factors to the discussion at hand. Due to the
energy threshold in Eqn.~(\ref{eqn:minE}) the WIMPs undergoing capture are
relatively energetic and have to transfer a substantial portion of their energy
to the nuclei. Therefore the recoil energy of the nuclei is large, and the process is very sensitive to the form factors used. If the Helm form factor overestimates the suppression for iodine at DAMA relative to iron for capture, it is possible that a relative factor of O(1) could result.

All together the uncertainties in astrophysical and nuclear inputs could lead 
to a factor of a few in the uncertainty of the capture rate of WIMPs in the 
Sun.  The DAMA/LIBRA modulation signal is much more sensitive to the velocity 
distribution of the Dark Matter particles, on the other hand.  This is because 
scatterings there only arise from particles in the tail of the velocity 
distribution.  It is encouraging that the results for capture presented here are so robust.  

\section{Capture vs. Annihilation in the Sun}
We now review the interplay between capture and annihilation in the Sun.
When the capture and annihilation rates are sufficiently large then equilibrium
will be reached between the two processes. 

Unless otherwise stated, we will assume that the capture and annihilation rates 
are sufficiently large so that the WIMPs are in equilibrium, and thus the annihilation rate 
is just one-half the capture rate: $\Gamma_{A} = C/2$. 


\subsection{Thermalization in Elastic Models}

A potentially major difference between the inelastic picture and the 
conventional picture is the question of what happens to WIMPs once they have scattered the first
time and been trapped in the Sun. For conventional WIMPs, repeated scatters
cause the WIMP to settle into thermal equilibrium in the center of the
Sun. In contrast, for an inelastic WIMP, if the kinetic energy is suitably low,
it is possible that no subsequent inelastic scatterings can take place. We must
consider what occurs under these circumstances as well.

Let us begin by assuming that there are {\it only} inelastic 
scatterings. In this case, the WIMP will proceed through the Sun, scattering 
off iron (which is abundant at the 10$^{-3}$ level), until such a point as it 
has insufficient kinetic energy to scatter off iron. There will be a range of 
orbits of varying ellipticity. Some will be circular, but most will be 
elliptical, and thus proceed more deeply into the Sun during their orbits. To 
determine how extensive this set
of orbits will be, we must consider how much kinetic energy a given WIMP will 
have in the interior of the Sun.

A WIMP starting at rest from the surface of the Sun will have a velocity
of approximately 1240 km/s in the center of the Sun.  This is more than enough
kinetic energy to scatter off iron for the parameters of interest. In fact, using the density
profile in the solar models of \cite{Bahcall:2004pz,PenaGaray:2008qe}, simply going from $R = 0.2R_{\bigodot}$ to the center will give enough energy to inelastically scatter off iron.
Thus, most WIMPs will be contained within this region.

We can now estimate whether the density in this region will be high enough to bring the system into equilibrium (i.e., the outgoing annihilation
rate equals the capture rate). If we assume a capture rate $C_{\bigodot}$, then (neglecting annihilation) the total number of particles in the Sun is $C_{\bigodot} \tau$ , where $\tau$ is the age of the Sun. If the WIMPs are captured within a radius $R=\epsilon R_{\bigodot}$, the present annihilation rate is then 
\begin{equation}
\Gamma_{ann} \approx \frac {(C_{\bigodot} \tau)^2}{ 2 \epsilon^{3} V_{\bigodot}} \sigma v.
\end{equation}
The annihilation will reach equilibrium with the capture if 
\begin{equation}
\sigma v \gsim \frac{ 2 \epsilon^{3} V_{\bigodot}}{ C_{\bigodot} \tau^{2}}
\approx 5 \times 10^{-28} cm^{3} s^{-1} \left( \frac{ 10^{24} s^{-1}}{ C_{\bigodot}} \right) \left(\frac{\epsilon}{.2}\right)^{3}.
\end{equation}
The thermal cross section needed to produce the correct relic abundance ($ 3 \times 10^{-26} cm^{2}$) is sufficent to put the WIMPs into equilbrium unless it decreases with velocity (e.g.~ is p--wave).  Should the WIMPs fail to reach equilibrium, there is an approximate overall suppression of the annihilation rate given by the LHS divided by the RHS.  Note that even in models where a p-wave cross section determines the relic abundance, there is also a subdominant  s-wave cross section.  This s-wave scattering could well be sufficient to put the WIMPs in equilibrium. 


Up to this point, we have considered particles with only inelastic scatterings,
but (sub-dominant) elastic scatterings are common in many explicit models of inelastic Dark Matter. For instance, mixed sneutrino and neutrino models have sizable Higgs mediated couplings  \cite{TuckerSmith:2001hy} ($\sigma_{n} \gsim 10^{-45}$ cm$^2$). Higgsinos--like models have loop-mediated contributions which can be important \cite{Hisano:2004pv} ($\sigma_n \gsim 10^{-48}$ cm$^2$). Generically, inelastic models will have cross sections which are $\sigma_n \sim (\delta/M_{WIMP})^{2} \sigma_{inelastic}  \approx   10^{-52}$ cm$^2$ \cite{TuckerSmith:2004jv}. So how large a cross section is needed to thermalize?

The total number of scatters in the Sun over the age of the Sun for a WIMP
would be
\begin{equation}
N_{scat} \approx
\frac{\bar{\rho}}{m_{p}} \sigma_{n} v \tau \approx 1 \times \left(\frac{\bar{\rho}}{1500 \, \rm{kg \cdot m^{-3} }}\right)
\left(\frac{\sigma_{n}}{2 \times 10^{-49} \, \rm{cm}^{2}}\right)
\left(\frac{v}{300 \, \rm{km \cdot s}^{-1}}\right)
\end{equation}
where $\bar{\rho}$ is the typical density seen by the WIMP and $m_p$ is the proton mass.  The solar density 
in the core is typically $1.5 \times 10^{5}$ kg $\cdot$ m$^{-3}$, roughly 100 
times larger than the mean solar density. To thermalize via scatterings off of  protons, the WIMP must scatter of order $M_{X}/m_p$ times.
Thus, a trapped 500 GeV WIMP scattering in the interior (with high densities) 
would need a cross section of approximately 10$^{-48}$ cm$^2$ to
thermalize.

Models with light mediators typically will not have this, but they often have
the Sommerfeld enhancement to boost annihilation rates to equilibrium. Models
that are not pure SU(2) doublets, but instead have some mixing with SU(2) singlets after EWSB generally have large enough Higgs couplings. Finally, models
which are pure SU(2) doublets - without Higgs couplings - generally have large
enough loop induced cross sections to thermalize them in the core, although this is marginal for WIMPs much heavier than 500 GeV.

Thus, we believe even in the inelastic case, it is reasonable to generally consider models that annihilate in equilibrium with their capture rates. We will make this assumption in what follows.

\section{Neutrino Rates}
Armed with a capture rate, we can now predict the high energy neutrino flux 
(and the resulting muon rates) on the Earth.  These rates will depend on the 
products of the WIMP annihilations.  We considered the final states of 
$WW$, $ZZ$, $b\bar{b}$ , $t \bar{t}$, $\tau \bar{\tau}$, $c \bar{c}$,  and light jets. We will not explicitly show separate results for $WW$ and $ZZ$ as they are very similar.

If WIMP annihilations are solely to charged leptons of the first two generations the annihilation products interact strongly enough in the Sun that they typically come to
rest before producing neutrinos.  The neutrinos from such decays are low
energy, and will not give an observable signal. A similar statement is true for particles that annihilate to first two generation/gluon jets.   The neutrino production comes from fragmentation to heavy quarks, and is very suppressed.   For our numerical results, we 
follow the work of Ref.~\cite{Cirelli:2005gh}, which builds on the work of 
Ref.~\cite{Jungman:1994jr}.

\subsection{Existing Limits}\label{sec:presentlimits}
Over the region of interest, the strongest bounds are placed by the 
Super-Kamiokande experiment~\cite{Desai} and the recent data from 
IceCube-22~\cite{IceCube22}.  At present, the limits are calculated assuming 
specific annihilation channels, which complicates the extraction of limits for 
a new model.  

To place limits on the model at hand using the data from Super-Kamiokande, we 
calculate the number of expected signal events at the detector, following the 
methods of Ref.~\cite{Delaunay:2008pc}.  We first compute the flux of neutrinos
from the Sun based on the appropriate annihilation channel.

We then propagate the neutrinos from the center of the Sun to the Earth, using 
the formula of Ref.~\cite{Cirelli:2005gh} which are available at 
Ref.~\cite{CirelliSite}.  These results for propagation are consistent with 
those of Ref.~\cite{EdsjoWimpSim}.   From this neutrino flux, we calculate the 
rate of muons at Super-Kamiokande by using the formula 
\begin{equation}
N_{evts}= \tau \int dE_\mu dE_\nu A_{eff}(E_{\mu}) \left[\left(\frac{d\sigma_{\nu p}}{dE_\mu} \rho_{p}  + \frac{d\sigma_{\nu n}}{dE_\mu} \rho_{n} 
\right) \frac{ d \Phi}{ d E_{\nu}} + (\nu \rightarrow \bar{\nu}) \right] R_{\mu}(E_{\mu}).
\label{eqn:evtsSK}
\end{equation}
The muon effective area $A_{eff}(E_{\mu}) = 1200$ m$^{2}$, and the livetime, $\tau$, is given by $\frac{1}{2} \times 1670$ days (the one-half is to account for night-time). The muon range, 
$R(E_{\mu})$, is approximately given by
\begin{equation}
R_{\mu} (E_\mu) = \frac{1}{\rho \beta} \log {\frac{ \alpha + \beta E_{\mu}}{
\alpha + \beta E_{thresh}} },
\end{equation}
where $\rho$ is the relevant density, $\alpha \approx 2.0$ MeV cm$^{2}$ g$^{-1}
$. $\beta$  varies depending on the material. For standard rock, $\beta \approx
 3 \times 10^{-6}$ cm$^{2}$ g$^{-1}$, whereas for water, $\beta \approx 4.2 \times 
10^{-6}$ cm$^{2}$ g$^{-1}$.  At Super--K, there can be conversions both in the 
nearby rock and in the water.  For simplicity, we will use the value $3 \times 
10^{-6}$ cm$^{2}$ g$^{-1}$, as rock usually dominates.  If we instead use the value 
for water the event rate increases by roughly 30$\%$. To compute the number 
events in Eqn.~(\ref{eqn:evtsSK}) we use the following neutrino-proton and
neutrino-neutron cross-sections \cite{nuxsec}:
\begin{eqnarray}
\frac{d\sigma_{\nu p}}{dE_\mu} &=& \frac{2}{\pi} m_{p} G_{F}^2 \left(a_{p \nu} + b_{p \nu} \frac{E_{\mu}^{2}}{E_{\nu}^2} \right),\\
\frac{d\sigma_{\bar{\nu} n}}{dE_\mu} &=& \frac{2}{\pi} m_{p} G_{F}^2 \left(a_{n \nu} + b_{n \nu} \frac{E_{\mu}^{2}}{E_{\nu}^2}\right).
\end{eqnarray}
Here $a_{p \nu}=0.15$, $b_{p \nu}=0.04$, $a_{n \nu}=.25$, $b_{n \nu}=0.06$, and the corresponding expressions for anti-neutrinos can be found by $a_{p \bar{\nu}}=b_{n\nu}$, $b_{p \bar{\nu}}=a_{n \nu}$, $a_{n \bar{\nu}}=b_{p \nu}$, $b_{n \bar{\nu}}=a_{p \nu}$.

The results from Eqn.~(\ref{eqn:evtsSK}) can then be compared to the data for upward going muons coming from the Sun as
reported in \cite{Desai}.  To calculate the number of observed ($N_{obs}$) and expected background ($N_{obs}^{bkgd}$) events
as a function of the WIMP mass we use data on upward going muons from Fig.5 of Ref.~\cite{Desai}.  The region of the sky
surrounding the sun used in the search varies as a function of WIMP mass.  We use Fig.8 of Ref.~\cite{Mori:1993tj} to specify
the angle about the sun that 
Sun as a function of the dark matter mass.  
through going muons is 
to Wimp annihilations in the Sun as a function 
\bea N_{exp}^{tot} = N_{obs}^{bkgd} + 0.9 N_{exp}^{signal} \label{Expevents} \eea where the factor of $0.9$ takes into
account the fact that only 90 \% of the signal is contained with the window angle in Fig.8 of Ref.~\cite{Mori:1993tj}.
Therefore the maximum allowed number of events is the value $N_{exp}^{signal}$ needed so that the cumulative poisson
distribution function is 10\%, for $N_{obs}$ number of observed events and $N_{exp}^{tot}$ number of expected events.  The
breaks in the curve around 90 and 225 GeV correspond to places where the size of the cone around the sun changes.  In that
cones size, there is a fluctuation in the number of events observed, which affects the limit.  Note that the optimal cone
size should actually depend on annihlation channel, as the neutrino spectra (and hence the correlation with direction) change
with final state.  However, the cone size in \cite{Mori:1993tj} is calculated instead with a ``breadbasket'' final
states\footnote{They take a final state of 80\% $b \bar{b}$, 10\% $c \bar{c}$ and 10\% $\tau^+ \tau^-$.}, so there is some
uncertainty on the exact position of the curves.

\begin{figure}[h!]
\includegraphics[width=3in]{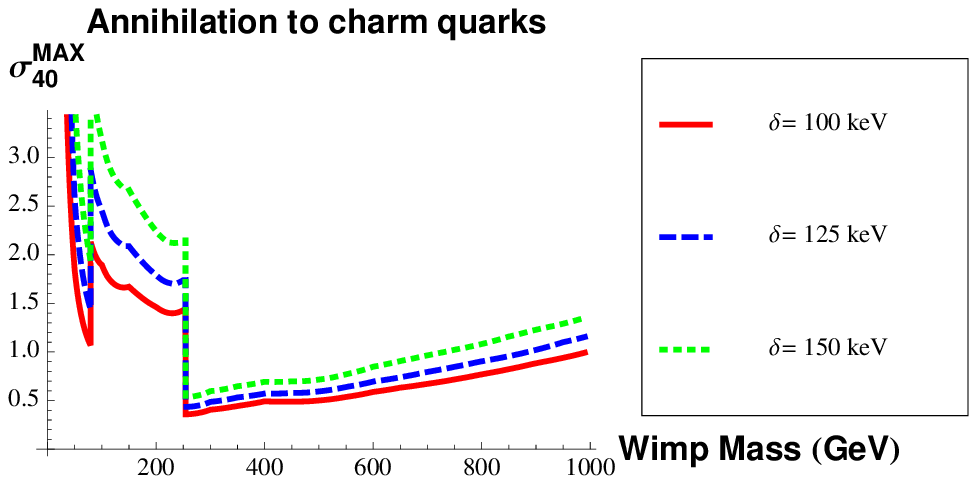}
\includegraphics[width=3in]{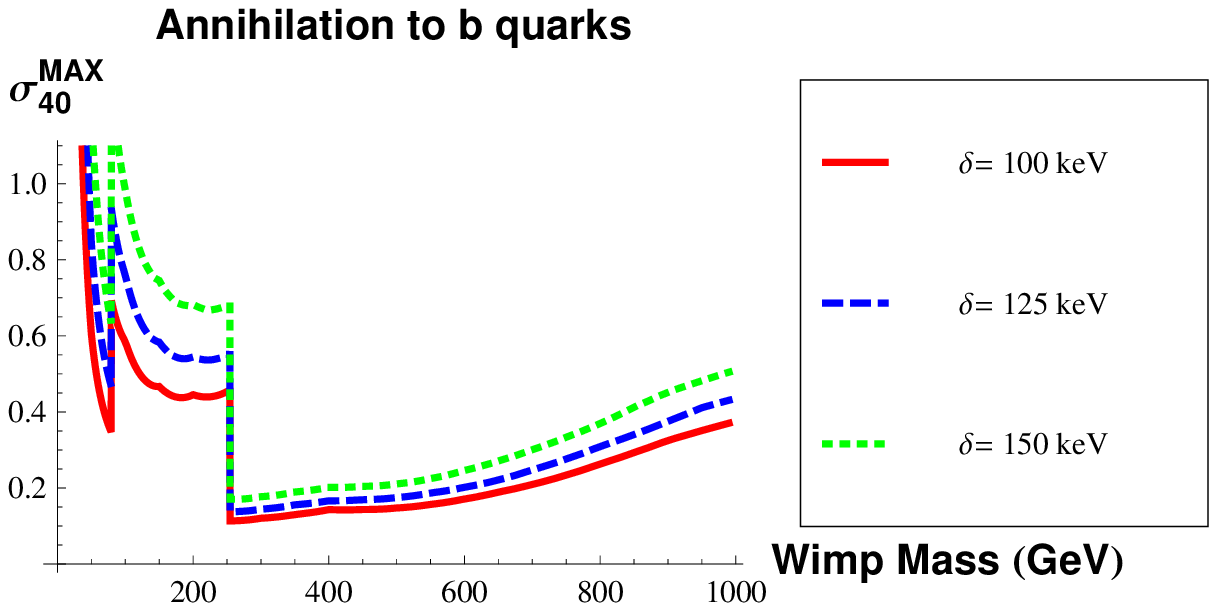}\\
\includegraphics[width=3in]{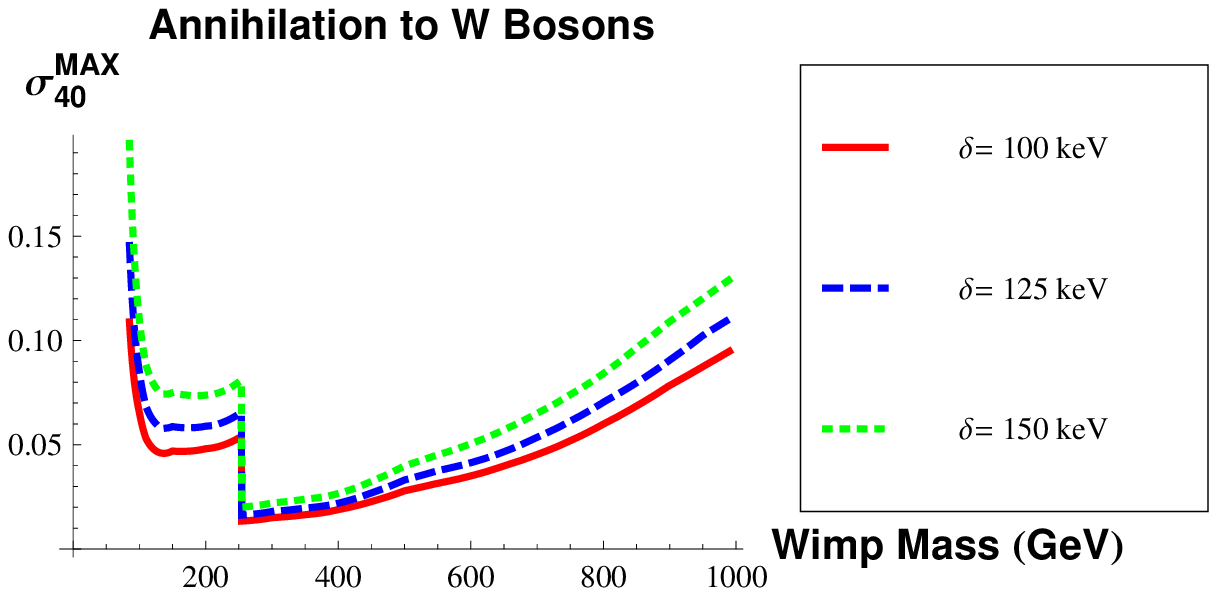}
\includegraphics[width=3in]{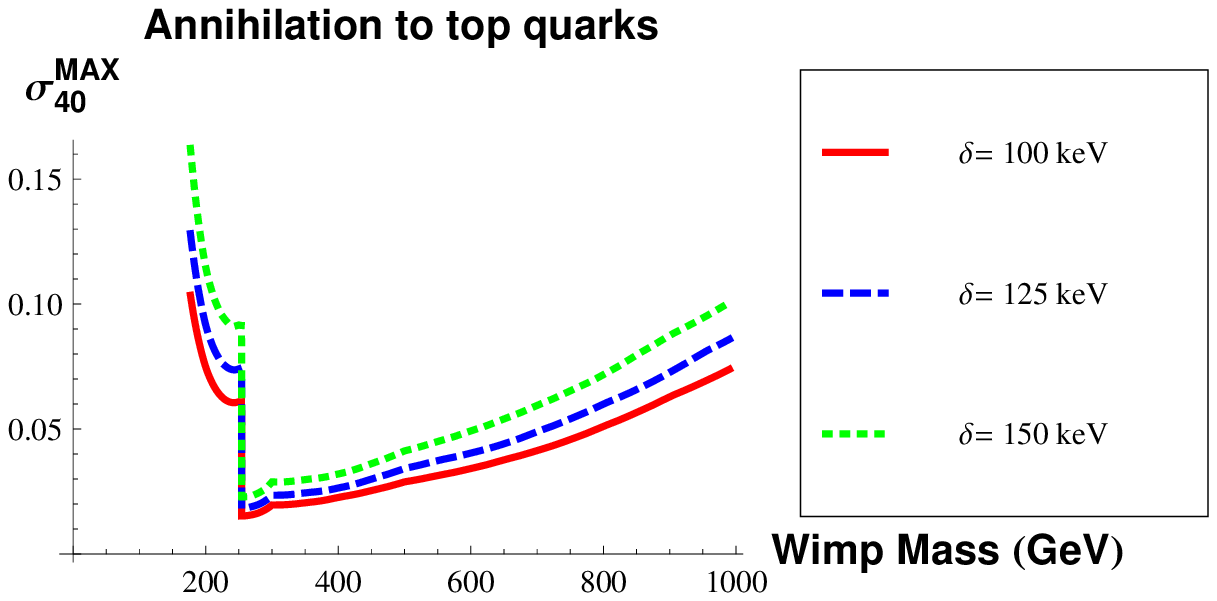}\\
\includegraphics[width=3in]{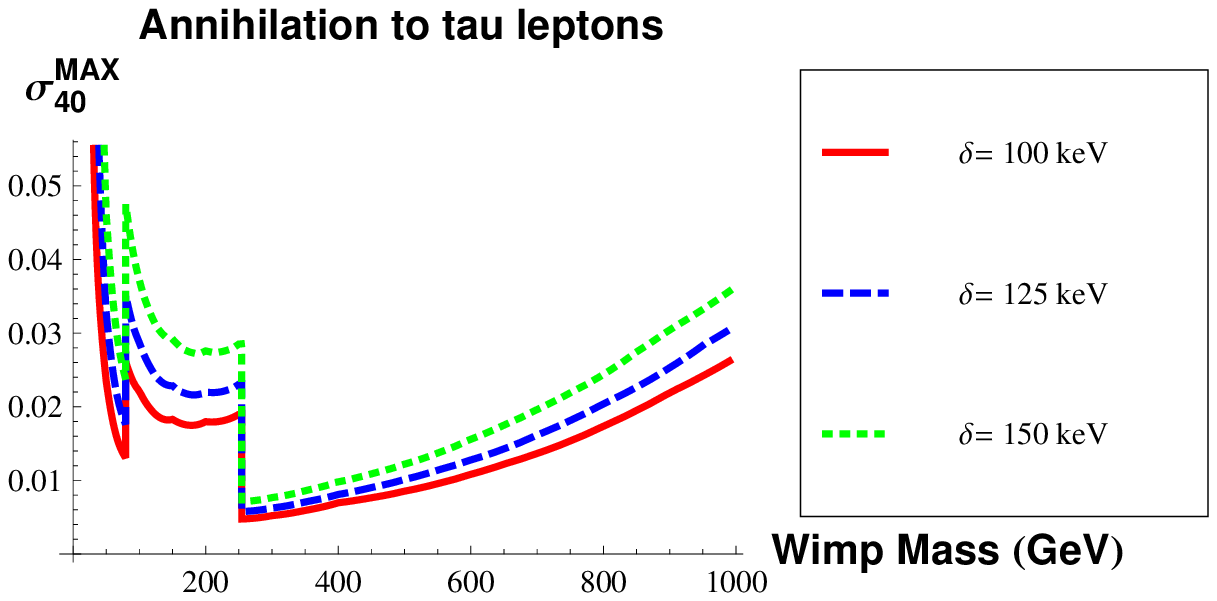}
\includegraphics[width=3in]{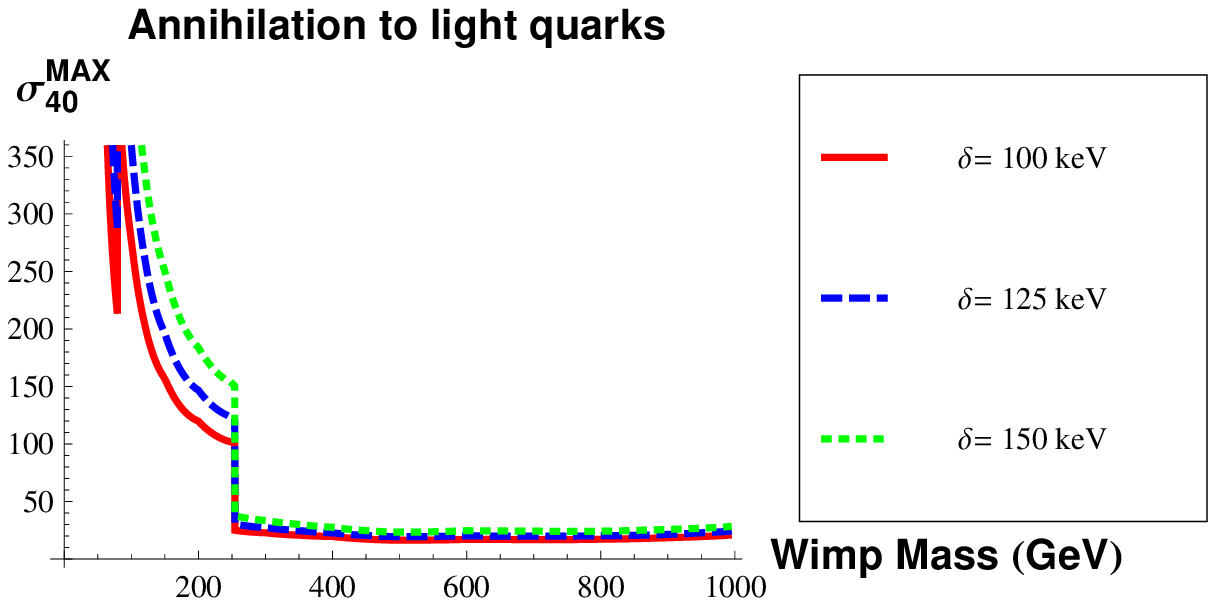}
\caption{We plot the maximum allowed cross section per nucleon (in units of of 
$10^{-40}$ cm$^{2}$) times the branching ratio to the given annihilation 
channel.  The curves are derived by imposing that the number of expected events (background + expected signal) is consistent with the number observed at the 90\% CL.      
For reference, cross sections consistent with the DAMA 
result are typically of the size $\sigma > 2 \times 10^{-40}$ cm$^{2}$, with 
$100 \mbox{ keV} <\delta <140$~keV .  We consider annihilations to several 
two-body Standard Model final states: W bosons, top quarks, charm quarks, 
b-quarks, light quarks, and tau leptons.  For large masses $>$ 250 GeV in the 
hard channel Ice-Cube22 stronger bounds (see text and Fig.~\ref{fig:Ice22}).}
\label{fig:SK}

\end{figure}

If the Dark Matter annihilates to a hard channel (e.g., W bosons, top quarks or
tau leptons) then the limits from Super--K are very constraining.  For cross 
sections consistent with the DAMA result, the typical event rates at Super--K 
would be too large by some two orders of magnitude (see Fig.~\ref{fig:SK}).  
If, instead the Dark Matter annihilates to a softer channel (bottom quarks or 
charm quarks), then the tension is lessened.

Nevertheless, it is fair to say that even in the case where the Dark Matter annihilates through a relatively soft channel such as charm (or more so with bottom quarks), there is tension with the existing limits from Super--Kamiokande.  At larger masses, the inelastic explanation for DAMA comes into tension with results from the CDMS  experiment \cite{JMR}.  However, N-body simulations have found significant structure in the high-velocity tails of velocity distributions \cite{Vogelsberger:2008qb}, and these may open the ranges of parameter space significantly when included \cite{kuhleninprogress}.

\begin{figure}[h]
\includegraphics[width=3in]{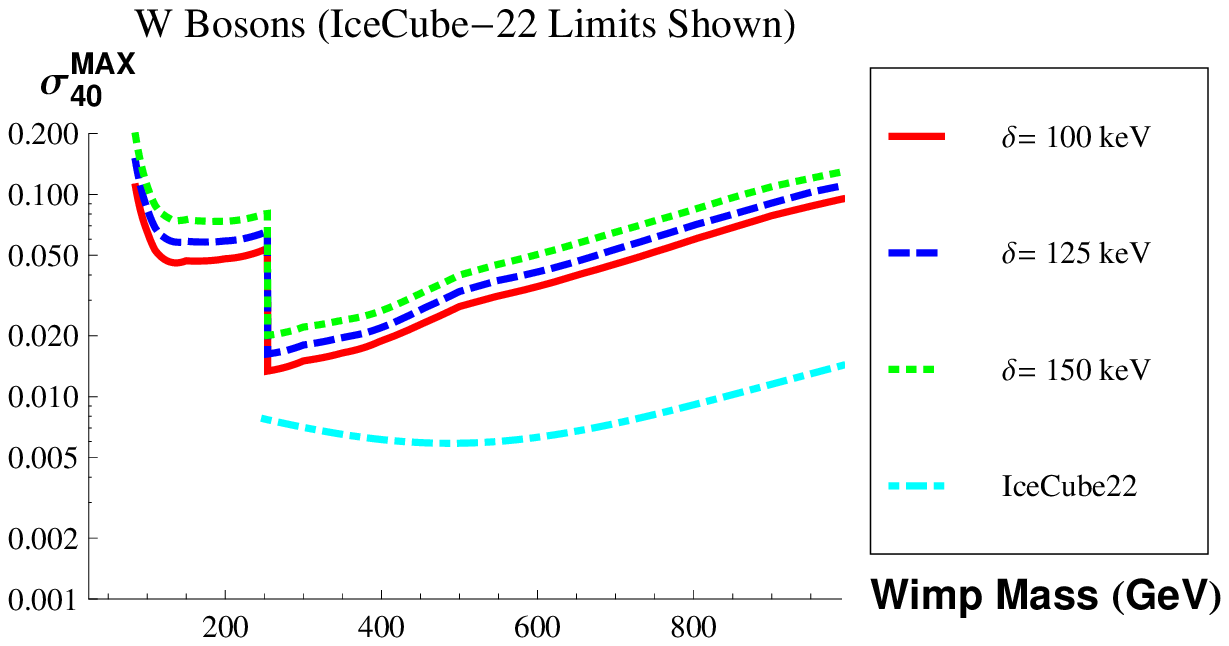}
\includegraphics[width=3in]{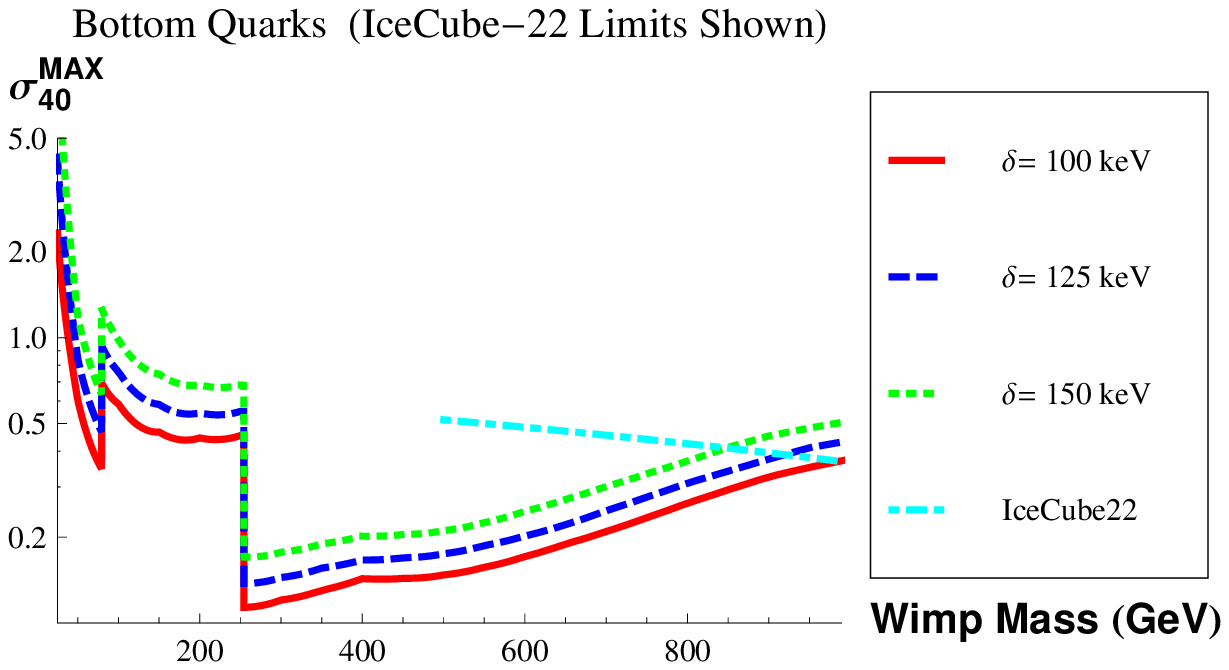}\\
\caption{As in the previous Figure, we plot the maximum allowed cross section 
per nucleon (in units of of $10^{-40}$ cm$^{2}$) times the branching ratio to a
given annihilation channel. The curves are derived by imposing that the number of expected events (background + expected signal) is consistent with the number observed at the 90\% CL.   
We also show limits extracted from the recent IceCube-22 results, which are 
(subtantially) stronger for the hard channel in the regime where they apply.}
\label{fig:Ice22}
\end{figure}

 It is important to note bounds from some channels (e.g.~ tau leptons) are much stronger than those from others (e.g.~ b quarks).  The result is that the dominant annihilation mode may not provide the most stringent bound on a given Dark Matter candidate.  The plots in Fig.~\ref{fig:SK} can be used to extract the maximum branching ratio to a given channel.   

Note, however, that at higher WIMP masses, (and especially for hard annihilation channels)
 the strongest limits come from the recent results of the 22-string run at 
IceCube.  In Ref.~\cite{IceCube22}, the collaboration quotes the maximum 
allowed values of the solar annihilation rate (= one-half capture rate in 
equilibrium) as a function of the neutralino mass for two different 
annihilation modes: $XX \rightarrow b \bar{b}$ (soft) and $XX \rightarrow W^{+}
 W^{-}$ (hard).  For the $WW$ annihilation mode, IceCube-22 excludes capture 
rates above $C_{\bigodot} > 1.2 \times 10^{22} s^{-1}$ ($3.2 \times 10^{21}$ s$^{-1}$) for $m= 500 (250)$ GeV.  No limits are quoted below 250 GeV.  
Comparing these rates with the capture rates of the previous section, we can 
see that these limits are very strong.  For the softer annihilation channel, 
the limits are weaker: $C_{\bigodot} < 2.8 \times 10^{23}$ s$^{-1}$ for $m=500$ 
GeV.  No constraints on lower masses for soft channels are given.  For 
reference, at 500 (250) GeV, a typical capture rate (for $\sigma_0 = 10^{-40}$ 
cm$^{2}$) is $8 (20) \times 10^{23}$ s$^{-1}$.  Thus for $b$ quarks, the IceCube 
constraint on the cross-section is roughly comparable to that of 
Super-Kamiokande for masses greater than 500 GeV (see Fig.~\ref{fig:Ice22}).  Because of the large energy threshold at IceCube-22, it is dangerous to extrapolate to lower masses, or to draw a strong conclusion about annihilation to charm quarks.  It is likely that IceCube should be able to probe these modes soon for higher masses (see next section).

For annihilation to $W$ bosons, IceCube-22 gives a bound that is nearly a factor of 4 stronger than Super-Kamiokande for masses greater than 250 GeV.  Again, it is dangerous to extrapolate to lower masses.  Something similar would be expected from other hard annihilation channels (tau leptons, top quarks, Z bosons).  It is clear that IceCube-22 places strong constraints on the hard annihilation channels.

\subsection{Future Telescopes}

We now turn to the sensitivity of future neutrino detection experiments to neutrino signals.  We concentrate on the case of the IceCube experiment.  We calculate both the expected number of events induced by neutrinos in the detectors, and the expected backgrounds that arise from the flux of atmospheric neutrinos.

For the atmospheric neutrino backgrounds, we use the data from 
Ref.~\cite{Honda:2006qj} to derive a power law fit, which takes the approximate form
\begin{eqnarray}
\Phi^{atm}(E_{\nu})  \approx 5.8 \times 10^{-2} E_{\nu}^{-3.14} \; \rm{cm}^{-2} \; GeV^{-1} \; sr^{-1} \; sec^{-1} \nonumber \\
\Phi^{atm}(E_{\nu})  \approx 5.6 \times 10^{-2} E_{\bar{\nu}}^{-3.20} \; \rm{cm}^{-2} \; GeV^{-1} \; sr^{-1} \; sec^{-1}
\end{eqnarray}
An experiment can focus only on the direction of the Sun to significantly reduce this background.  We integrate over a 3${^\circ}$ region for IceCube.

\begin{figure}
\includegraphics[width=3in]{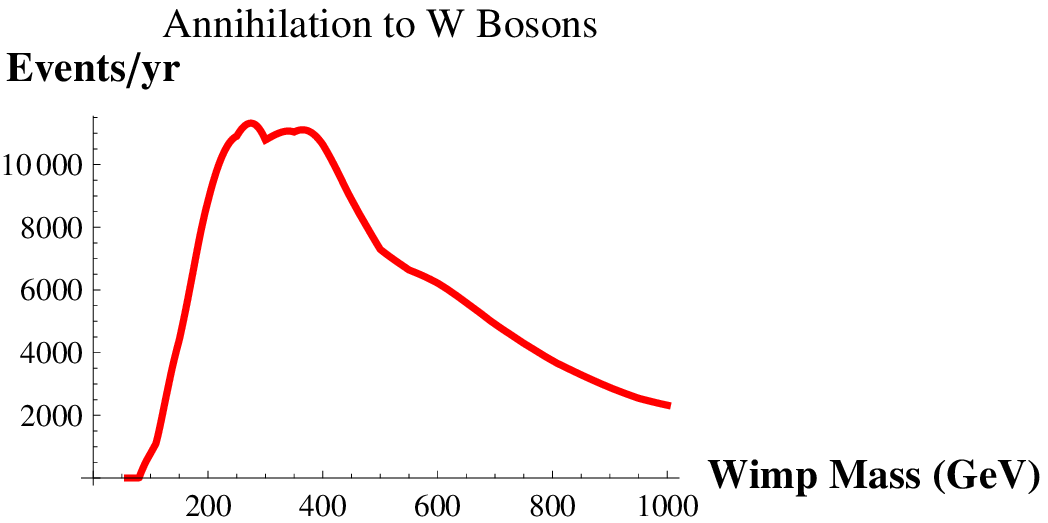}
\includegraphics[width=3in]{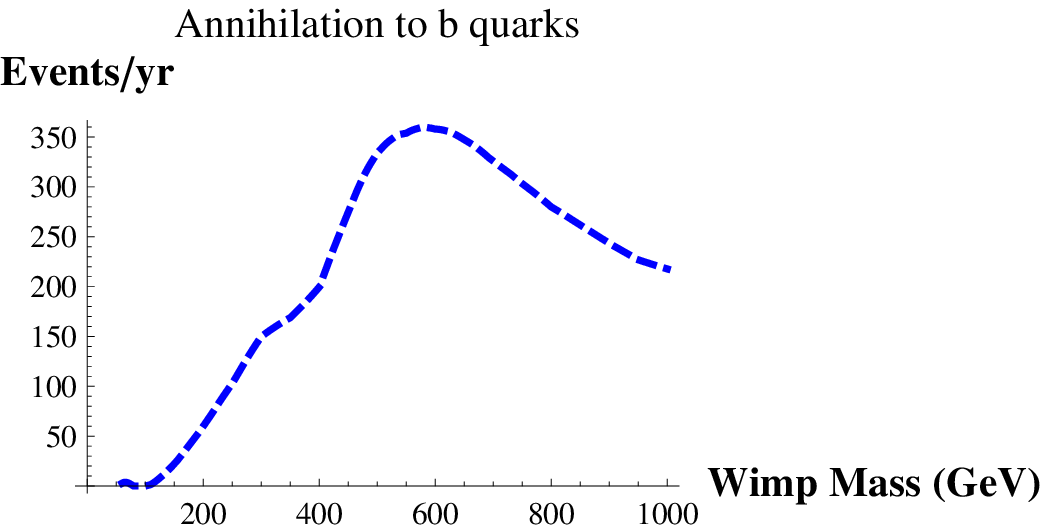}
\caption{The number of signal events expected in one year of running at IceCube, assuming a cross section of $\sigma_{0}=10^{-40}$ cm$^{2}$ per nucleon and a $\delta = 125$ keV and annihilation to W bosons (left) and b quarks (right).  Note such a large cross section is excluded in the case of the $W$ boson, see Figs.~\ref{fig:SK} and \ref{fig:Ice22}.  }
\label{fig:IceCubeEvts}
\end{figure}

To calculate the actual number of events at 
IceCube, we again use Eqn.~(\ref{eqn:evtsSK}).
For the purposes of determining the energy loss in ice, we take $\alpha \approx
2.0$ MeV cm$^{2}$ g$^{-1}$ and $\beta \approx 4.2 \times 10^{-6}$ cm$^{2}$ g$^{-1}$. We take a threshold energy of $E_{thresh}= 50$ GeV, which is somewhat optimistic but may be attainable, especially with the planned inclusion of the of the DeepCore array \cite{Resconi:2008fe}.
We take the effective area $A_{eff}(E_{\mu})$ for IceCube from Ref.~\cite{GonzalezGarcia:2005xw}.    We find approximately 40 background events arising form atmospheric neutrinos within a $3^{\circ}$ window.  The corresponding number of signal events, normalized to a cross section $\sigma_{0} = 10^{-40}$ cm$^{2}$ are shown in Fig.~\ref{fig:IceCubeEvts}.

If nature has chosen the past of inelastic Dark Matter, perhaps the mostly likely reason that it has gone undetected thus far is annihilation proceeds to modes that provide soft neutrinos.  Thus, it will be important for IceCube (and other future experiments such as ANTARES) to try and push their energy threshold as low as possible.  

 \section{Conclusions}
Inelastic dark matter provides an exciting proposal to reconcile the results of DAMA/LIBRA with other direct detection experiments. However, its large cross section makes detection through neutrino telescopes a particularly important constraint. We have found that these experiments can place strong limits on these models, and so it is important to clearly state what those constraints are.

Under the assumption that inelastic dark matter s-wave annihilates with a thermal cross section, it certainly appears that for hard annihilation channels (e.g.~ tau leptons, W bosons, monochromatic neutrinos) such a scenario is excluded.  For other cases where the annihilation products are softer sources of neutrinos (charm quarks or b quarks), the answer is less clear. Annihilation to charm, as well as other light quarks seems safe, at least within astrophysical uncertainties  (though the tension in the charm channel is large at larger masses). Annihilation into b quarks seems excluded at the factor of two level (and more for higher masses). When including the variety of uncertainties, both from astrophysics as well as other issues such as nuclear form factors this is borderline. It is important to note, however, that annihilations into Higgs bosons generally include a component of $\tau$ as well, which even as subdominant contributions are often the dominant limit. In particular when considering the IceCube-22 limits, particles above $\sim 250$ GeV are strongly constrained.

We should note that these limits are possibly evaded in elastic models when the annihilation rate is p-wave suppressed, and equilibrium between annihilation and capture has not yet been achieved. 

Importantly, models in which the dark matter annihilates into new, light force carriers that dominantly decay into $e^+e^-$, $\mu^+ \mu^-$ and $\pi^+ \pi^-$ appear safe from these constraints, because muons and pions stop before decaying and producing neutrinos.

Another possibility to evade these bounds is that the Dark Matter might dominantly annihilate to large multiplicity final states.  In this case, the neutrino energies are degraded, as the energy is shared amongst a large number of final decay products, and limits might easily be evaded.  This could occur naturally in light mediator models, where one might have $XX \rightarrow \phi \phi \rightarrow 4b$, or models where the Higgs boson dominantly decays via pseudo-scalars~\cite{WeinerDermisekChang}, in which case $XX \rightarrow h h \rightarrow 4 a \rightarrow 8b$.  If the light states ($\phi$ or $a$) are allowed to decay to $\tau$ leptons or directly to neutrinos, then tension may still exist.  A detailed examination of such decays and related model building is left for future work\cite{WorkInProgress}.  Because the bounds on the channels with energetic neutrinos are so much stronger, it is possible (or even likely) that a sub-dominant decay mode with hard neutrinos could provide the strongest constraint for models of this type.  This is also of potential relevance if Higgs boson decays are involved and both $\tau$ leptons and $b$-quarks are potentially present.

In the case where Dark Matter is not its own anti-particle, and possesses a conserved quantum number, the Dark Matter abundance is due to a small excess of Dark Matter over anti--Dark Matter.  Then captured Dark Matter may not annihilate and there will be no signal.  Such a scenario is incompatible with potential annihilation signals observed by PAMELA and FGST.

Ultimately, the space of models to explain DAMA through inelastic scattering is still large, but the space is strongly constrained by these neutrino telescopes. Should future direct detection bear out the presence of inelastic WIMPs, particularly at higher masses, these null results should allow us to distinguish among a variety of candidates.

 \begin{acknowledgments}
 We would like to thank Patrick Fox, Dan Hooper, Michele Papucci, and Chris Savage for useful discussions, and Itay Yavin for many helpful discussions and informing us of his work\cite{Itay}. The authors also thank Carlos Pe\~na-Garay for providing us with detailed information about solar elemental distributions. The work of A.~Menon and A.~Pierce is supported under DOE Grant \#DE-FG02-95ER40899.  The work of A.~Pierce is also supported by NSF CAREER Grant NSF-PHY-0743315.  The work of R.~Morris and N.~Weiner is supported by NSF CAREER grant PHY-0449818 and DOE OJI grant \#DE-FG02-06ER41417.\end{acknowledgments}

\bibliography{IDM}
\bibliographystyle{apsrev}
\end{document}